\documentclass[aps,prl,twocolumn,showpacs,superscriptaddress,groupedaddress]{revtex4}  
\usepackage{graphicx}  
\usepackage{dcolumn}   
\usepackage{bm}        
\usepackage{amssymb}   
\usepackage{subfigure}
\usepackage{amsmath}

\begin{document}

\title{Losses-based test of wave-particle duality with Mach-Zehnder interferometers}

\author{Chen~Yang}
\affiliation{The State Key Laboratory of Quantum Optics and Quantum Optics Devices, Institute of Opto-Electronics, Shanxi University, Taiyuan 030006, China}
\author{Aiai~Jia}
\affiliation{The State Key Laboratory of Quantum Optics and Quantum Optics Devices, Institute of Opto-Electronics, Shanxi University, Taiyuan 030006, China}
\affiliation{Beijing Computational Science Research Centre, Beijing 100084, China}
\author{Xue~Deng}
\affiliation{The State Key Laboratory of Quantum Optics and Quantum Optics Devices, Institute of Opto-Electronics, Shanxi University, Taiyuan 030006, China}
\author{Yuchi~Zhang}
\affiliation{The State Key Laboratory of Quantum Optics and Quantum Optics Devices, Institute of Opto-Electronics, Shanxi University, Taiyuan 030006, China}
\affiliation{College of Physics and Electronics Engineering, Shanxi University, Taiyuan 030006, Shanxi Province, China}
\author{Gang~Li}
\affiliation{The State Key Laboratory of Quantum Optics and Quantum Optics Devices, Institute of Opto-Electronics, Shanxi University, Taiyuan 030006, China}
\author{Shiyao~Zhu}
\email[]{syzhu@csrc.ac.cn}
\affiliation{The State Key Laboratory of Quantum Optics and Quantum Optics Devices, Institute of Opto-Electronics, Shanxi University, Taiyuan 030006, China}
\affiliation{Beijing Computational Science Research Centre, Beijing 100084, China}
\author{Tiancai~Zhang}
\email[]{tczhang@sxu.edu.cn}
\affiliation{The State Key Laboratory of Quantum Optics and Quantum Optics Devices, Institute of Opto-Electronics, Shanxi University, Taiyuan 030006, China}

%
%

\vskip 0.25cm
\date{\today}

\begin{abstract}
Wave-particle duality of photons with losses in the Mach-Zehnder interferometer (MZI) is investigated experimentally and theoretically. The experiment is done with the standard MZI with the beam splitter or the beam merger being continuously varied. The losses are deliberately introduced either inside the MZI (the two arms between the beam splitter and beam mergers) or outside the MZI (after the beam merger). It is proved that the unbalanced losses have great influence on the predictability $P$ (particle nature) and visibility $V$ (wave nature). For the former case the duality inequality holds while for  the later the duality inequality is ``violated''. We get $P^2+V^2>1$. This ``violation'' could be eliminated in principle by switching the two paths and detectors and then averaging the results. The observed results can be exactly explained theoretically. The experiment is done with coherent beam, instead of single photons, and we have proved that they are exactly equivalent in duality experiment with MZI.
\end{abstract}

\maketitle

The non-intuitive quantum behavior of wave-particle duality plays the center role in quantum mechanics \cite{1} and this weird property of a particle has been extensively studied for decades. Wooter and Zurek first studied quantitatively the wave-particle duality \cite{2} in 1979. In 1986, Glauber discussed the relation between which-way information and the fringe contrast in Young's double-pinhole interference experiment with single incident photon and he got the relation of $P^2+V^2=1$, where $P$ is the difference between the probabilities of the photon goes to the two pinholes and $V$ is the fringe contrast \cite{3}. He wrote: \emph{``It is only when we have no inkling of which way the photon went that we can see fringes with strong contrast, $V=1$. Complementarity, in short, has won again. The incident photon can behave either as a particle or a wave, but it never exhibits both extremes of behavior at once. What one usually sees in neither the one behavior nor the other and the equations describe a whole continuum of possible compromises.''} Later the inequality of $P^2+V^2\leq1$ was experimentally tested by Greenberger and Yasin \cite{4} and Mandel \cite{5} and theoretically obtained by Jaeger et al. \cite{6} and Englert \cite{7}. The inequality of $P^2+V^2\leq1$ characterizes the duality of the photon. In the experiment with Mach-Zehnder interferometer (MZI), $P$ is the predictability of the photon and $V$ is the visibility of the interference behind the MZI. The predictability is experimentally obtained by the difference between the two probabilities ($w_1$ and $w_2$) the photon goes one way and the other, i.e., 
\begin{equation}
\label{eq:1}
P=\lvert w_1-w_2 \rvert.
\end{equation}
The visibility experimentally obtained by the interference fringes,
\begin{equation}
\label{eq:2}
V=\left|\frac{w_{\textnormal{max}}-w_{\textnormal{min}}}{w_{\textnormal{max}}+w_{\textnormal{min}}}\right|,
\end{equation}
with $w_{\textnormal{max(min)}}$ being the maximum (minimum) detection probability at one of the two detectors.Experimental tests were carried out  by S. D\"urr et al. with a single-photon interferometer \cite{8} and a device of four-beam-interference \cite{9}. In the past few years, there has been renewed interest in the wave-particle duality investigation based on MZI \cite{10, 11}. The second beam splitter of the MZI could be set as ``on'' or ``off'', or even in a status of quantum superposition of ``on'' (``present'') or ``off'' (``absent''), which is called as quantum beam splitter (QBS). Some people have suggested that this quantum MZI allows one to investigate both the wave nature and the particle nature of photons \cite{12}. This has inspired some experimental investigations \cite{13,14,15,16,17,18} with different physical systems. The experiment with QBS has proved that the wave-particle duality relation is still satisfied \cite{13,14,15,16,17}. In 2013, an experiment by using the QBS came out with an interesting result with $P^2+V^2>1$, which seems to violate the duality and this result was explained as the interference between the particle and wave behaviors \cite{18}. The discussion of wave-particle relation has also extended to the multi-photon with high order wave-particle duality \cite{19,20} and multiple internal degrees \cite{21}. It has been indicated that $P^2+V^2\leq1$ is valid for a state of any number of photons.

In all above discussions on the duality, it is assumed that there is no loss on the paths and the efficiencies of the detectors are perfect or ignored them. However, the photon losses are inevitable in any experiment. Our study shows that the influence of the photon losses on the duality need to be considered \cite{22}. Here we report the experimental study about the influence of the losses on the duality. Our experimental results, which are consistent with our theoretical prediction, tell us that photon losses have truly distinct features. The experiment is done with two configurations of the MZI with a variable beam splitter (VBS) either in the input port (beam splitter) or output port (beam merger). The photon losses are deliberately introduced either inside the two arms of the MZI or outside the arms. Here, we emphasis that the experiment is carried out with a coherent beam as the input and the detectors are intensity detectors. In the supplementary material, we have theoretically shown that the result with coherent input is exactly the same as the result with a single photon input. The duality relation is also valid for coherent beam input. The results of the experiment shed new light on this fundamental relation of wave-particle duality.

\begin{figure}
\includegraphics[scale=0.4]{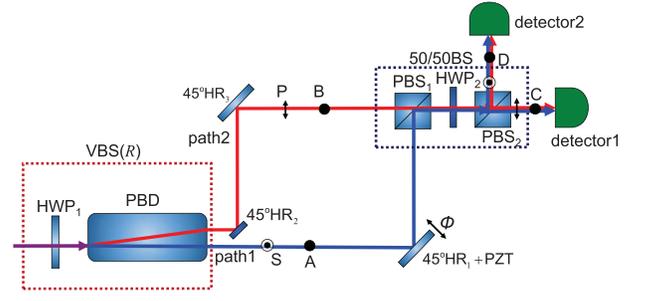}
\caption{\label{fig:1} Experimental setup for wave-particle relation test with photon losses. VBS: variable beam splitter; HWP: half wave plate; PBD: polarization beam displacer; PZT: piezoelectric transducer. S and P stand for the polarizations corresponding to the blue and red line for path 1 and path 2, respectively. Points A, B, C, D are the places where the filters are inserted.}
\end{figure}

\begin{figure}
\centering
\includegraphics[scale=0.35]{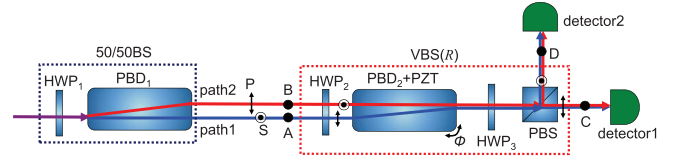}
\caption{\label{fig:4} Second configuration of the experiment. The beam splitter of the MZI is a fixed 50/50 BS and the beam merger is a VBS. Other elements are the same as Fig.~\ref{fig:1}. }
\end{figure}

The two configurations are shown in figure ~\ref{fig:1} and figure ~\ref{fig:4}, respectively. In the first configuration (Fig.~\ref{fig:1}), the beam splitter is a VBS and the beam merger is a fixed 50/50 beam splitter. In the second configuration (Fig.~\ref{fig:4}) the beam splitter is a fixed 50/50 beam splitter, while the beam merger is a VBS. These two configurations were proved to be equivalent if the MZI only has one beam on one input port (another one is vacuum) and there is no any losses \cite{23}.

The experimental setup for the first configuration is shown in Fig.~\ref{fig:1}. A coherent beam provided by a frequency-stabilized cw single-longitudinal-mode diode laser (Toptica DL100$@$894) is sending to the VBS (in the red dashed rectangular). The VBS is formed by a half wave plate (HWP$_1$) and a polarization beam displacer (PBD), which separate the input beam into two orthogonally polarized beams with high extinction ratio. After path 1 and path 2 the two beams are recombined by the 50/50 beam merger (in the blue dashed rectangular) which is formed by two polarization beam splitters (PBS) and HWP$_2$. The outputs of the MZI are detected by two detectors in the end.

The 50/50 BS can be realized by setting the angle between axis of HWP$_2$ and PBS$_2$ to $22.5^{\circ}$. By carefully aligning the two paths, the maximum visibility of MZI can be reached to $99.2\%$. In this MZI we independently introduce two kinds of losses: one inside the MZI (point A and B), and another outside the MZI (point C and D). In each case, in order to deliberately introduce the unbalanced losses, two absorptive neutral density filters with different absorptions are inserted in the optical paths either at point A and B or point C and D. The absorptions of the filter at point A (or C) are calibrated as $L_1=72.7\%$, and another at point B (or D) it is $L_2=39.6\%$. By rotating the HWP$_1$, the splitting ratio of VBS can be changed continuously. The visibility ($V$) (Eq. (A1) in the supplementary material) and the predictability ($P$) (Eq. (A2) in the supplementary material) are measured under different VBS settings with different splitting ratio $R$. The relation of $P^2+V^2$ versus the splitting ratio $R$ is thus obtained (Eqs. (A12) and (A15) in the supplementary material).

It should be noted that the quantum efficiencies of detector 1 ($Q_1$) and detector 2 ($Q_1$) and the measured visibility $V_0$ are not perfect. In the case of losses inside the MZI, Eqs. (A7), (A11) and (A12) in the supplementary material become:
\begin{equation}
\label{eq:3}
V=\frac{2\sqrt{R(1-R)(1-L_1)(1-L_2)}}{(1-R)(1-L_1)+R(1-L_2)}\times{V_0},
\end{equation}
\begin{equation}
\label{eq:4}
P=\frac{\left|(1-R)(1-L_1)/Q_1-R(1-L_2)/Q_2\right|}{(1-R)(1-L_1)/Q_1+R(1-L_2)/Q_2},
\end{equation}
\begin{eqnarray}
\label{eq:5}
\begin{aligned}
P^2+V^2=&\frac{4R(1-R)(1-L_1)(1-L_2)}{[(1-R)(1-L_1)+R(1-L_2)]^2}\times{V_0^2}\\
&+\frac{\left|(1-R)(1-L_1)/Q_1-R(1-L_2)/Q_2\right|^2}{[(1-R)(1-L_1)/Q_1+R(1-L_2)/Q_2]^2}.
\end{aligned}
\end{eqnarray}
When the losses are outside the MZI, from Eqs. (A14), (A11) and (A15) in the supplementary material we get:
\begin{equation}
\label{eq:6}
V=2\sqrt{R(1-R)}\times{V_0},
\end{equation}
\begin{equation}
\label{eq:7}
P=\frac{\left|(1-R)(1-L_1)/Q_1-R(1-L_2)/Q_2\right|}{(1-R)(1-L_1)/Q_1+R(1-L_2)/Q_2},
\end{equation}
\begin{equation}
\label{eq:8}
\begin{aligned}
P^2+V^2=&\frac{\left|(1-R)(1-L_1)/Q_1-R(1-L_2)/Q_2\right|^2}{[(1-R)(1-L_1)/Q_1+R(1-L_2)/Q_2]^2}\\
&+4R(1-R)\times{V_0^2}.
\end{aligned}
\end{equation}

\begin{figure}
\centering
\includegraphics[scale=0.25]{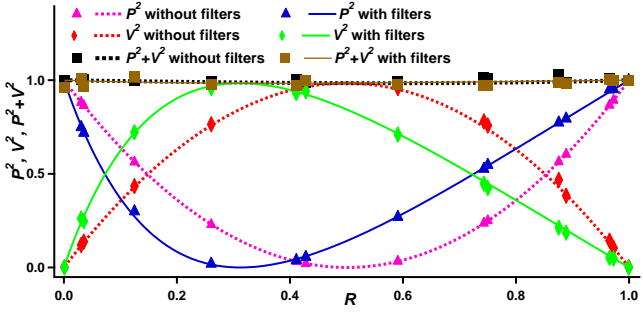}
\caption{\label{fig:2}  The result for losses inside the MZI. The pink triangles, the red diamonds and the black squares stand for the measured values of $P^2, V^2, P^2+V^2$ without filters ($L_1=0, L_2=0$), while the blue triangles, the green diamonds and the brown squares stand for the measured values of $P^2, V^2, P^2+V^2$ with filters ($L_1=72.7\%, L_2=39.6\%$) in path 1 and path 2. All the curves are the fittings according to Eqs.~(\ref{eq:3}), ~(\ref{eq:4}) and ~(\ref{eq:5}) with $Q_1=90.4\%, Q_1=90.8\%$ and $V_0=99.2\%$.}
\end{figure}

Figure~\ref{fig:2} shows the experimental results. The pink triangles, the red diamonds and the black squares stand for the measured $P^2$, $V^2$, $P^2+V^2$ without filters in the paths, respectively, while the blue triangles, the green diamonds and the brown squares are data of measured $P^2$, $V^2$, $P^2+V^2$ with filters. The curves in Fig.~\ref{fig:2} are the corresponding fittings by Eqs.~(\ref{eq:3}), ~(\ref{eq:4}) and ~(\ref{eq:5}). It is clear that the measured result is in good agreement with the theory. Here we have used the measured parameters: $Q_1=90.4\%$, $Q_2=90.8\%$ and $V_0=99.2\%$.

From the result in Fig.~\ref{fig:2} we can see that when the losses are inside the MZI, both $P^2$ and $V^2$ are not symmetrical any more, but the wave-particle relation $P^2+V^2 =1$ still holds. Losses inside MZI have influences on the visibility and predictability, but have no effect on the wave-particle duality relation.
\begin{figure}
\centering
\includegraphics[scale=0.25]{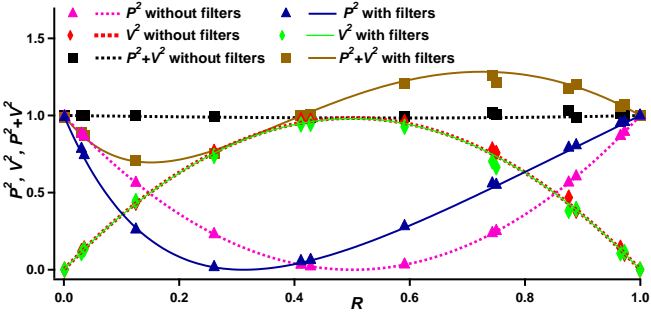}
\caption{\label{fig:3} The results for losses outside the MZI. All the marks and parameters are same as in Fig.~\ref{fig:2}, except that the filters are inserted outside the MZI at point C and D, and the theoretical fittings are based on Eqs.~(\ref{eq:6}), ~(\ref{eq:7}) and ~(\ref{eq:8}).}
\end{figure}

Figure~\ref{fig:3} shows the results for the case when losses are outside the MZI. In this case the two filters are inserted just before the two detectors at point C and D. All the rest remains unchanged. In this situation, $P^2$ is not symmetrical any more and $V^2$ does not change. It is obvious that the measured  wave-particle inequality $P^2+V^2\leq1$ does not hold any anymore. Here, the losses have no effect on the visibility but have an influence on the predictability. In fact, when $R$ is between 0.43 to 1, the ``violation'' of the wave-particle inequality is observed. At $R=0.74$, we get $P^2+V^2=1.26$, which is greater than 1. This ``violation'' comes from the fact that part of the photons is lost and the measured predictability is not the originally defined predictability \cite{7} anymore.

Now let's turn to the second configuration (Fig.~\ref{fig:4}). We fix the beam splitter of the MZI as 50/50 BS and the beam merger is a VBS and then just repeat the experimental procedure as before.

Consider the real quantum efficiencies of detectors and the visibility, in the condition of losses inside the MZI, Eqs. (B5), (B8) and (B9) in the supplementary material become:
\begin{equation}
\label{eq:9}
V_1=\frac{2\sqrt{R(1-R)(1-L_1)(1-L_2)}}{(1-R)(1-L_1)+R(1-L_2)}\times{V_0},
\end{equation}
\begin{equation}
\label{eq:10}
V_2=\frac{2\sqrt{R(1-R)(1-L_1)(1-L_2)}}{(1-R)(1-L_2)+R(1-L_2)}\times{V_0},
\end{equation}
\begin{eqnarray}
\label{eq:11}
\begin{aligned}
P=&\frac{1}{2}\bigg[\frac{\left|(1-R)(1-L_2)/Q_1-R(1-L_2)/Q_2\right|}{(1-R)(1-L_2)/Q_1+R(1-L_2)/Q_2}\\
 &+\frac{\left|(1-R)(1-L_1)/Q_1-R(1-L_1)/Q_2\right|}{(1-R)(1-L_1)/Q_1+R(1-L_1)/Q_2}\bigg],
\end{aligned}
\end{eqnarray}
\begin{equation}
\label{eq:12}
P^2+V_1^2=P^2+\frac{4R(1-R)(1-L_1)(1-L_2)\times{V_0^2}}{[(1-R)(1-L_1)+R(1-L_2)]^2},
\end{equation}
\begin{equation}
\label{eq:13}
P^2+V_2^2=P^2+\frac{4R(1-R)(1-L_1)(1-L_2)\times{V_0^2}}{[(1-R)(1-L_2)+R(1-L_2)]^2}.
\end{equation}
And in the condition of losses outside the MZI, Eqs. (B11), (B14), (B15) in the supplementary material become:
\begin{equation}
\label{eq:14}
V=2\sqrt{R(1-R)}\times{V_0},
\end{equation}
\begin{eqnarray}
\label{eq:15}
\begin{aligned}
P=&\frac{1}{2}\bigg[\frac{\left|(1-R)(1-L_2)/Q_2-R(1-L_1)/Q_1\right|}{(1-R)(1-L_2)/Q_2+R(1-L_1)/Q_1}\\
 &+\frac{\left|(1-R)(1-L_1)/Q_1-R(1-L_2)/Q_2\right|}{(1-R)(1-L_1)/Q_1+R(1-L_2)/Q_2}\bigg],
\end{aligned}
\end{eqnarray}
\begin{eqnarray}
\label{eq:16}
\begin{aligned}
P^2+V^2=&\frac{1}{4}\bigg[\frac{\left|(1-R)(1-L_2)/Q_2-R(1-L_1)/Q_1\right|}{(1-R)(1-L_2)/Q_2+R(1-L_1)/Q_1}\\
&+\frac{\left|(1-R)(1-L_1)/Q_1-R(1-L_2)/Q_2\right|}{(1-R)(1-L_1)/Q_1+R(1-L_2)/Q_2}\bigg]^2\\
&+4R(1-R)\times{V_0^2}.
\end{aligned}
\end{eqnarray}

\begin{figure}
\centering
\subfigure[visibility detected by detector 1 ($V_1$)]
{\includegraphics[scale=0.25]{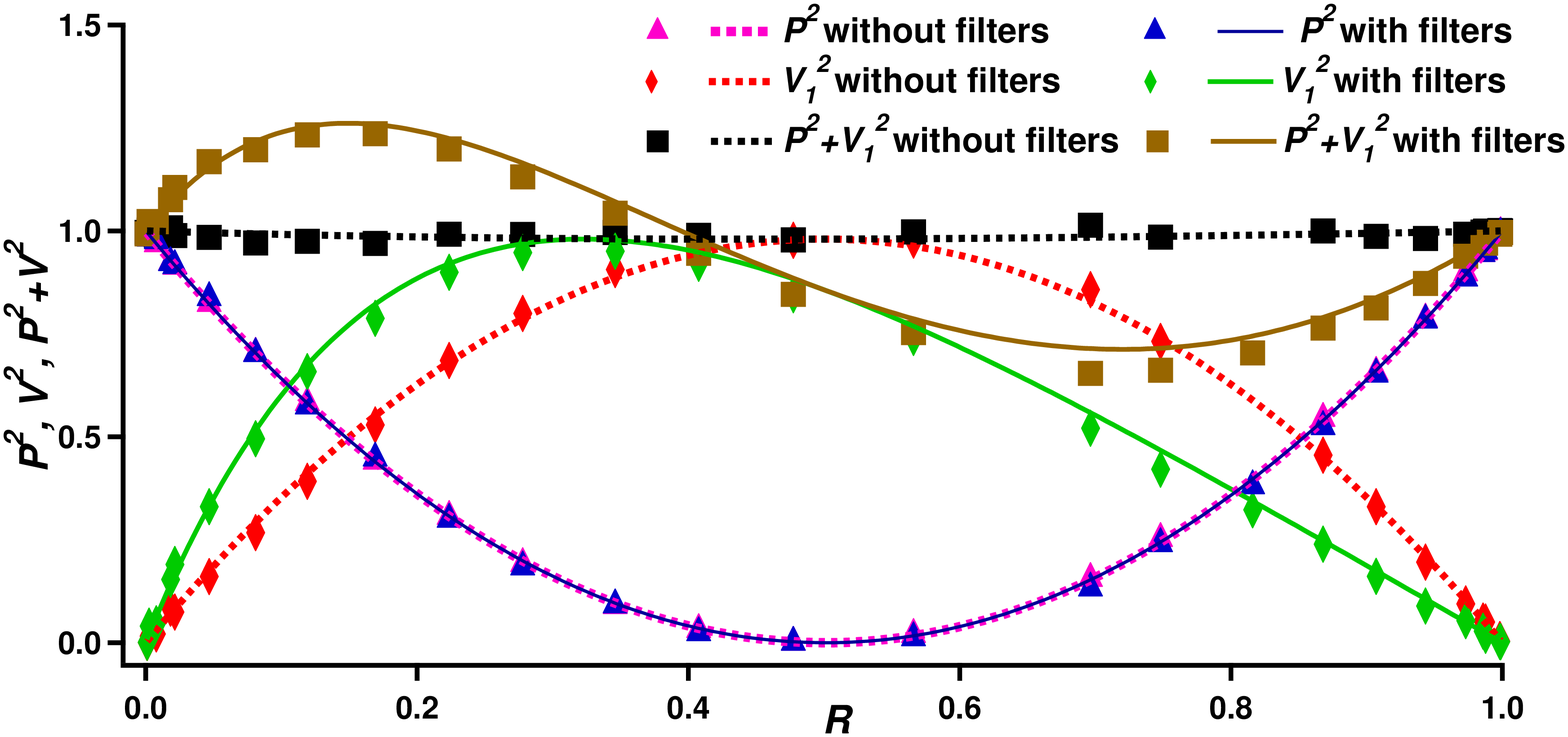}
\label{fig:5a}}

\subfigure[visibility detected by detector 2 ($V_2$).]
{\includegraphics[scale=0.25]{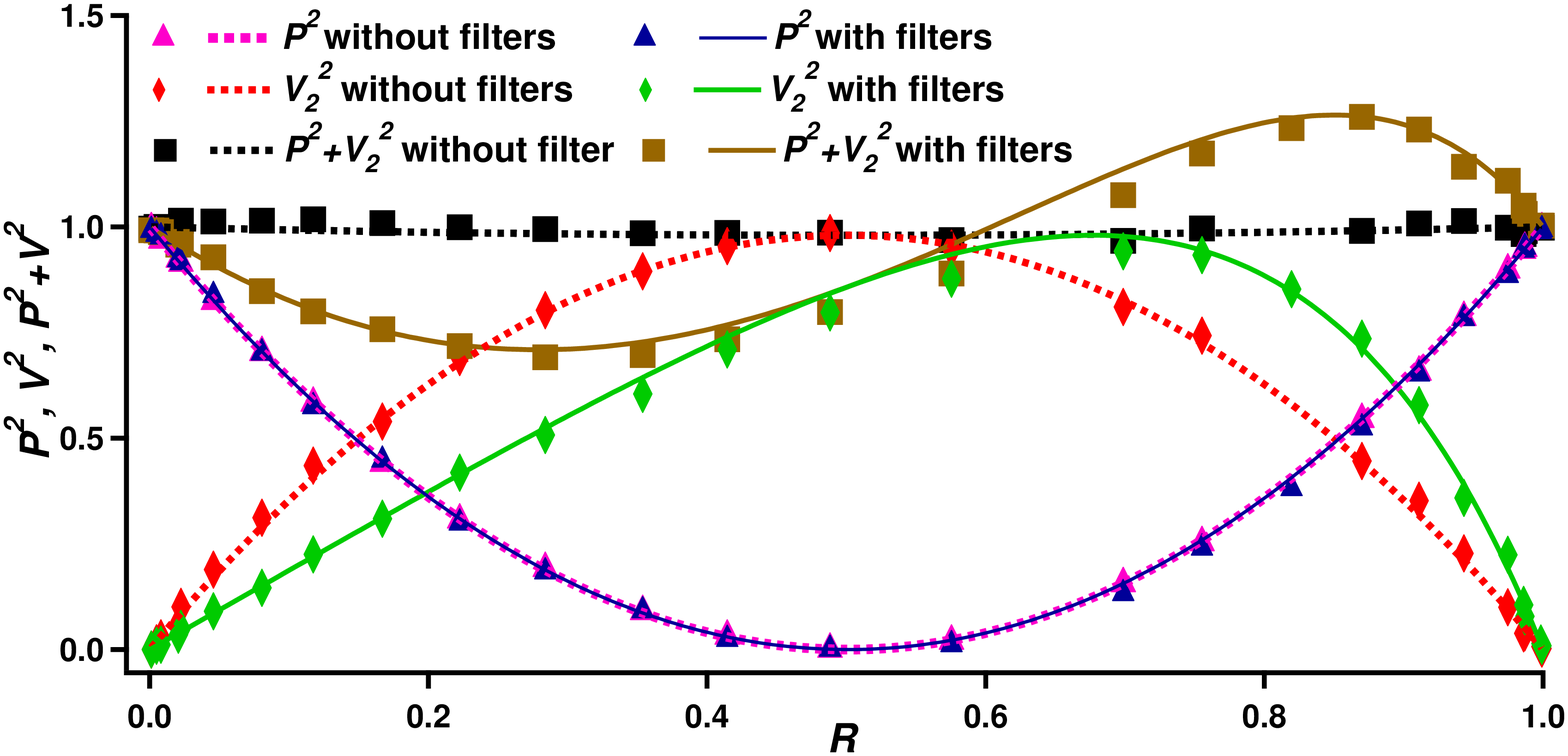}
\label{fig:5b}}
\caption{Experiment result based on the second configuration with losses inside the MZI. All the marks are the same as in Fig.~\ref{fig:2}. (a) is data from detector 1 and (b) is data from detector 2. The curves are fittings by using Eqs.~(\ref{eq:9})-(\ref{eq:13}). The losses of the filters: $L_1=76.4\%$, $L_2=50.5\%$ in path 1 and path 2, respectively, and $V_0=99.0\%$.} \label{fig:5}
\end{figure}
Figure~\ref{fig:5} shows the results and the corresponding fittings by Eqs.~(\ref{eq:9})-(\ref{eq:13}) for the case that losses are inside the MZI. It is clear that the measured results are again in agreement with the theory. Here we have used the measured data: $Q_1=90.4\%$, $Q_2=90.8\%$ and $V_0=99.0\%$. The visibilities measured by detector 1 (see Fig.~\ref{fig:5a}) and 2 (see Fig.~\ref{fig:5b}) are not the same, and these are different from what we observed in the first configuration. The data of $V^2$ is not symmetrical. The value of $P^2+V^2$ thus changes as the merging ratio $R$ varies from 0 to 1. That is to say, the losses make influence on the wavelike behavior. The visibilities measured by detector 1 and 2 are different. However they have no effect on the predictability. The wave-particle inequality is again ``violated''.

\begin{figure}
\centering
\includegraphics[scale=0.25]{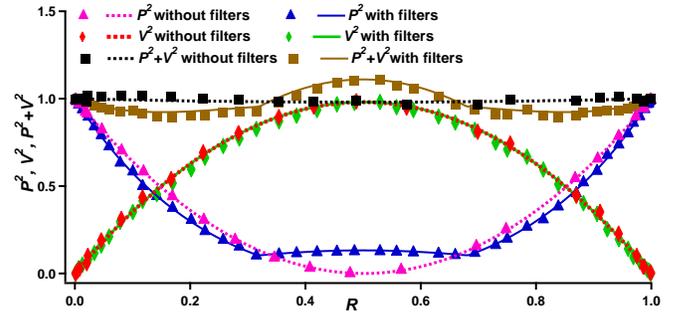}
\caption{\label{fig:6}Experiment result based on the second configuration with losses outside the MZI. All the marks and parameters are the same as in Fig.~\ref{fig:5}, except that $L_1=73.5\%$, $L_2=43.4\%$ and theoretical fittings are from Eqs.~(\ref{eq:14}), ~(\ref{eq:15}) and ~(\ref{eq:16}).}
\end{figure}
The result with the losses outside the MZI is illustrated in Fig.~\ref{fig:6}. We see that both $P^2$ and $V^2$ are still symmetrical when $R$ varies from 0 to 1. $V^2$ is independent on the losses while $P^2$ has two ``turning points'' at $R=0.68$ and $0.32$, which result in a ``straw hat type'' of the $P^2+V^2$. This weird feature comes from the fact that the particlelike behavior depends on the predictabilities detected by detector 1 and detector 2, and the minimum values of $P_1$ and $P_2$ are not at the same point (Eq. (B14) in the supplementary material). That is to say, the losses make influence on the predictability, but have no effect on the visibility and the wave-particle duality relation along with $R$ shows distinct features compare to the results from the first configuration shown in Fig.~\ref{fig:3}.

We would like to emphasize that the ``violation'' is due to the unbalanced losses introduced in the two paths either inside or outside the MZI. So, this ``violation'' could in principle be eliminated either by switching the input beam from one port to another one or switching the two optical paths, including the detectors, and then averaging the results before and after. The ``violations'' have nothing to do with the re-normalization of the beam light but simply the unbalanced losses of the MZI. We actually have renormalized the beam since part of the light is not measured. The maximum ``violation'' is $P^2+V^2=2$, which corresponds to the cases that the losses are outside the MZI with $R=0.5$ and one of the beam is blocked ($L_1=1$).

In conclusion, we have demonstrated a losses-based wave-particle duality test by a standard MZI. The experiment is done with coherent beam, instead of using single photons or single qubits. We have proved that the exact same result will be obtained in the duality experiment with MZI either by using single photon input or by coherent beam input. Two experiment configurations are finished and for each one the losses inside or outside the MZI are investigated. All the observed features of $P^2+V^2$ along with the splitting ratio (or merging ratio) $R$ can be well explained by simply inducing the unbalanced losses, including the observed ``violation'' of the wave-particle inequality $P^2+V^2>1$. The ``violation'' comes from the fact that when losses appear in MZI, part of light is lost and the measured predictability and visibility are not the original definitions given by Englert \cite{7}, and the measured wave behavior and particle behavior are not exactly the behaviors of the photon just coming out from the first beam splitter. The inevitable photon losses in real experiment are not trivial. The losses do have great influence on the interpretation of the MZI data and play an important role in testing the wave-particle inequality. The results of the experiment will shed new light on this fundamental wave-particle duality relation.

%
We would like to thank Chuanfeng Li for his reading the manuscript and helpful suggestions. This work was supported by National Basic Research Program of China (2012CB921601, 2012CB921603) and National Natural Science Foundation of China (Grants Nos. 11125418, 91336107, 61275210, 61227902, 61121064).
%

\end{document}